\documentclass[twocolumn,groupaddress,amsmath,aps,longbibliography,nofootinbib]{revtex4-1}

\usepackage{hyperref}
\usepackage{graphicx}
\usepackage{amsmath,amsfonts}
\usepackage{dcolumn}
\usepackage{bm}
\usepackage{color}
\usepackage{multirow}
\usepackage{float}
\usepackage{url}
\usepackage{physics}
\usepackage{subfigure}
\usepackage[utf8x]{inputenc}
\usepackage[T1]{fontenc}

\begin{document}

\title{Electronic structure of CeIr$_3$ superconductor: DMFT studies}

\author{Sylwia Gutowska} 
 \email{gutowska@agh.edu.pl}
\author{Bartlomiej Wiendlocha} 
\address{Faculty of Physics and Applied Computer Science,
AGH University of Science and Technology, Aleja Mickiewicza 30, 30-059 Krakow, Poland}

\date{\today}

\begin{abstract}
We present the band structure of CeIr$_3$ superconductor calculated within the dynamical mean field theory (DMFT). 
Standard GGA and GGA+U methods fail to reproduce the experimental electronic specific heat coefficient $\gamma_{\rm expt.}$ due to the underestimated density of states at the Fermi level $N(E_F)$ followed by an overestimated strength of electron-phonon coupling (EPC) calculated as a renormalization of $\gamma_{\rm expt.}$.
The DMFT study shows a strong hybridization of $4f$ states of Ce with $5d$-states of Ir, which leads to the larger $N(E_F)$ giving the correct $\gamma_{\rm expt.}$ with a moderate EPC in agreement with the experimental data. 
\end{abstract}

\maketitle

\section{Introduction}
The compounds of rare earth metals and transitional metals are intensively studied due to multiple physical phenomena such as heavy fermionic character, Kondo effect or RKKY interactions \cite{cerium-compounds2} and the competition of magnetic order and superconductivity. The interplay of $f$-states of Cerium and $d$- or $p$-states of transitional metals varies among all these materials and often may lead to a mixed or intermediate valency of cerium, which results in various magnetic properties \cite{cerium-compounds}. Additionally, while mostly $f$-states are localized, the band-like character of $f$-state is also possible, as shown in case of CeIr$_2$ \cite{ceir2}.

Regarding the superconductivity, the Ce $\alpha$-phase superconducts under the pressure of 50\,GPa \cite{ce-SC} and to the best knowledge of authors, the superconductivity in family of binary compounds of cerium and transitional metals under ambient pressure is known only in a few cases, such as CeRu$_2$ ($T_c$=6\,K) \cite{ceru2-sc}, CeCo$_2$ (1.5\,K) \cite{ceco2-sc}, CeIr$_5$ (1.85\,K) and CeIr$_3$ (2.5\,K) \cite{ceir5-sc}.  Additionally, while most of Ce$M_3$ compounds crystallize in cubic structure, the CeIr$_3$ and CeCo$_3$ are the only two which crystallize in the $R\overline{3}m$ crystal structure. 

As shown in previous works \cite{ceir3,ceir3-1}, CeIr$_3$ is a nonmagnetic, type-II superconductor with $T_c$=2.5\,K and moderately-strong electron-phonon coupling (EPC), with a generally conventional BCS-type behavior seen e.g. in the heat capacity measurements~\cite{ceir3}.
The EPC constant $\lambda_{\rm ep}$, estimated based on the $T_c$ and McMillan formula\cite{mcmillan}, is $\lambda_{\rm ep} = 0.65$.
On the other hand, 
two-gap superconductivity was suggested based on the anisotropy of the upper magnetic critical field \cite{ceir3-2}, whereas $s$-wave superconductivity competing with weak spin fluctuations was determined based on the $\mu$SR studies \cite{ceir3-musr}.

Our previous band structure calculations \cite{ceir3}, done using GGA~\cite{pbe} and GGA+U \cite{gga_u} approach (both including spin-orbit coupling) showed that $d$-states of Ir dominate near the Fermi level of CeIr$_3$. $4f$ states or Ce were found to be of a nonmagnetic itinerant character, and integration of the Ce $f$-like density of states gave the $f$-state filling of 1.05 for GGA and 0.20 for GGA+U (U = 2.5 eV), pushing the $f$-states above the Fermi level in the latter case.
However, none of this calculations were able to reproduce the electronic heat capacity, showing that the electronic structure near the Fermi energy $E_F$ is not properly described. The measured value of the Sommerfeld coefficient was $\gamma_{\rm expt.} = 25.1$, whereas the computed were 10.16 (GGA) and 5.71  (GGA+U), all in $\frac{\rm{mJ}}{\rm{mol}\cdot\rm{K}^2}$. 
Assuming that the renormalization of the electronic specific heat is mostly due to electron-phonon interactions characterized by the EPC constant $\lambda_{\rm ep}$, i.e. $\gamma_{\rm expt.} = \gamma_{\rm calc.}(1+\lambda_{\rm ep})$, we arrive at the much overestimated values of $\lambda_{\rm ep}$ = 1.47 (GGA) or 3.40 (GGA+U), inconsistent with the magnitude of $T_c$. The problem arrives from the too small value of the density of states at the Fermi level, $N(E_F)$.  That is why the GGA+U, which pushes the $f$-states above $E_F$, gives way too high estimate of $\lambda_{\rm ep}$.
he reason for this is that GGA+U works well for compounds with localized $f$ states such as EuPd$_3$ \cite{eupd3} or EuIr$_2$Si$_2$ \cite{euir2si2} and fails when $f$ states are hybridized and well dispersed \cite{dmft0}.

To overcome this difficulty, in this work we present the electronic structure calculated with the  DMFT framework \cite{dmft,dmft2}, which allows for a better treatment of the correlated $4f$-states of cerium, especially when they are hybridized, i.e. the Coulomb interaction energy is close to the electronic bandwidth \cite{dmft0}, as it is in CeIr$_3$. 
To our knowledge, it is the first band structure of a binary Ce-$M$ ($M$ - transitional metal) superconductor calculated within the DMFT method.
The electronic band structure has been determined using WIEN2k package~\cite{wien2k}  and DFT+eDMFT (embedded dynamical mean field theory) software \cite{dmft2}. Unit cell parameters and atomic position optimized in previous work \cite{ceir3} were used in the self-consistent cycle, with the grid of 21x21x21 $\bf{k}$-points. In DMFT calculations, the continuous-time quantum Monte Carlo (CTQMC) impurity solver has been used with the exact double counting scheme \cite{exact}, which allows not to assume the starting occupancy of the $f$-states. The Slater parameter used for calculation of the Slater integrals (i.e. the Coulomb term) \cite{edmft2}, is set to $F_0=6$\,eV in agreement with $U$ value calculated with Madsen-Novak's method~\cite{novak} (see below) and with the value commonly used for Ce. The electronic temperature~\cite{dmft2} was set to 116\,K, the value applied before in case of cerium \cite{edmft}\footnote{Lower temperatures would require more Matsubara frequency points in calculations, making them very time-consuming \cite{edmft})}. Additional calculations performed at the temperature of 300\,K resulted in the same electronic structure, thus we may assume that our results correspond to the ground state. Additionally, experimental data analyzed in the previous work~\cite{ceir3} revealed no thermal occupancy of the $4f$ state.

\section{Results}

We start with revoking the GGA and GGA+U results, which are next compared to the DMFT studies.

\begin{figure}
    \centering
    \includegraphics{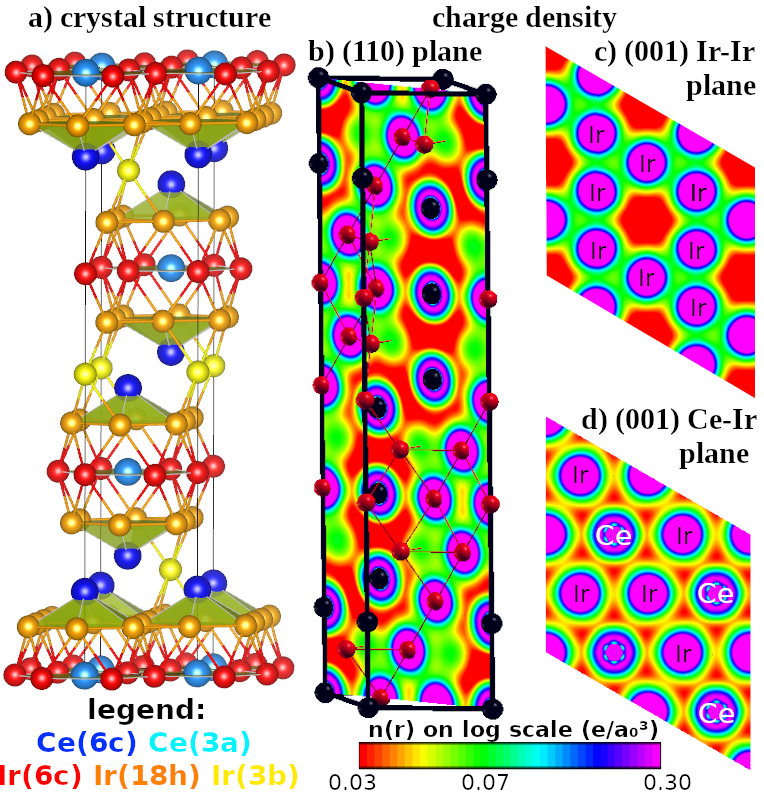}
    \caption{Crystal structure of CeIr$_3$ (a), electronic charge density shown in (110) plane (b) and two (001) planes consisted of Ir(18h) atoms (c) and both Ir(6c) and Ce(3a) atoms (d).}
    \label{fig:cell}
\end{figure}

\paragraph{GGA and GGA+U}
Most of the members of Ce$M_3$ family ($M$ - transition metal) crystallize in a simple cubic $Pm\overline{3}m$, Cu$_3$Au-type structure, where all nearest-neighbor interatomic distances are equal.
Only the CeIr$_3$ and CeCo$_3$ compounds, both with seven valence $d$ electrons, form in the rhombohedral $R$-3$m$ phase. 
In this structure, two Ir atoms are distorted towards each other, reducing their atomic distance and forming a close-packed tetrahedron, and in total there are three nonequivalent Ir atoms (18h, 6c, 3b sites). The Ce atoms occupy two nonequivalent positions, 3a and 6c.

The conventional cell of this structure is presented in Fig. \ref{fig:cell}. It consists of metallic layers of Ir(18h) and layers of both Ir(6c) and Ce(3a) atoms, while Ce(6c) and Ir(3b) are located between the layers. As shown in Fig. \ref{fig:cell}, the charge delocalization connected to the metallic character of the bonds is much stronger in Ir-Ir layer ($d_{\rm{Ir-Ir}}=2.64$\,\AA), where atoms are closer to each other than in case of Ir-Ce layer ($d_{\rm{Ce-Ir}}=3.06$\,\AA). 

\begin{table}[b]
\caption{The electronic and superconducting properties of CeIr$_3$ calculated with GGA, GGA+U and DMFT: occupation of $4f$ states $n(4f)$, total and partial DOS at $E_F$ (eV$^{-1}$ per f.u.), bandstructure  $\gamma_{\rm calc.}$ and experimental $\gamma_{\rm expt.}$ Sommerfeld coefficients ($\frac{\text{mJ}}{\text{mol}\text{K}^2}$), EPC constant from Eq.(1) without and with the assumed spin fluctuations ($\lambda_{\gamma}$ and $\lambda_{\gamma,ep}=\lambda_{\gamma}-\lambda_{sf}$ respectively)  and calculated from McMillan's formula ($\lambda_{ep}(T_c)$)}
\label{tab:results}
\begin{ruledtabular}
\begin{tabular}{llll}
& GGA &	GGA+U&	DMFT\\
\hline
$n(4f)$& 	1.05&	0.2&	0.67\\
$N(E_F)$& 	4.31&	2.42&	5.83\\
$N_{3Ir}(E_F)$ &  2.32&	2.02&	3.17\\
$N_{Ce}(E_F)$ &	1.51	&0.36&	2.66\\
$\gamma_{\rm calc.}$	&10.16&	5.71&	13.74\\
\hline
$\gamma_{\rm expt.}$	&\multicolumn{3}{c}{25.1}	\\	
\hline
$\lambda_{\gamma}$&	1.47&	3.4&	0.83\\
$\lambda_{\gamma,ep}$& ---&--- &	0.77\\
\hline
$\lambda_{ep}(T_c)$ &	\multicolumn{3}{c}{0.65}\\
\end{tabular}
\end{ruledtabular}
\end{table}

\begin{figure}
    \centering
    \includegraphics[width=.5\textwidth]{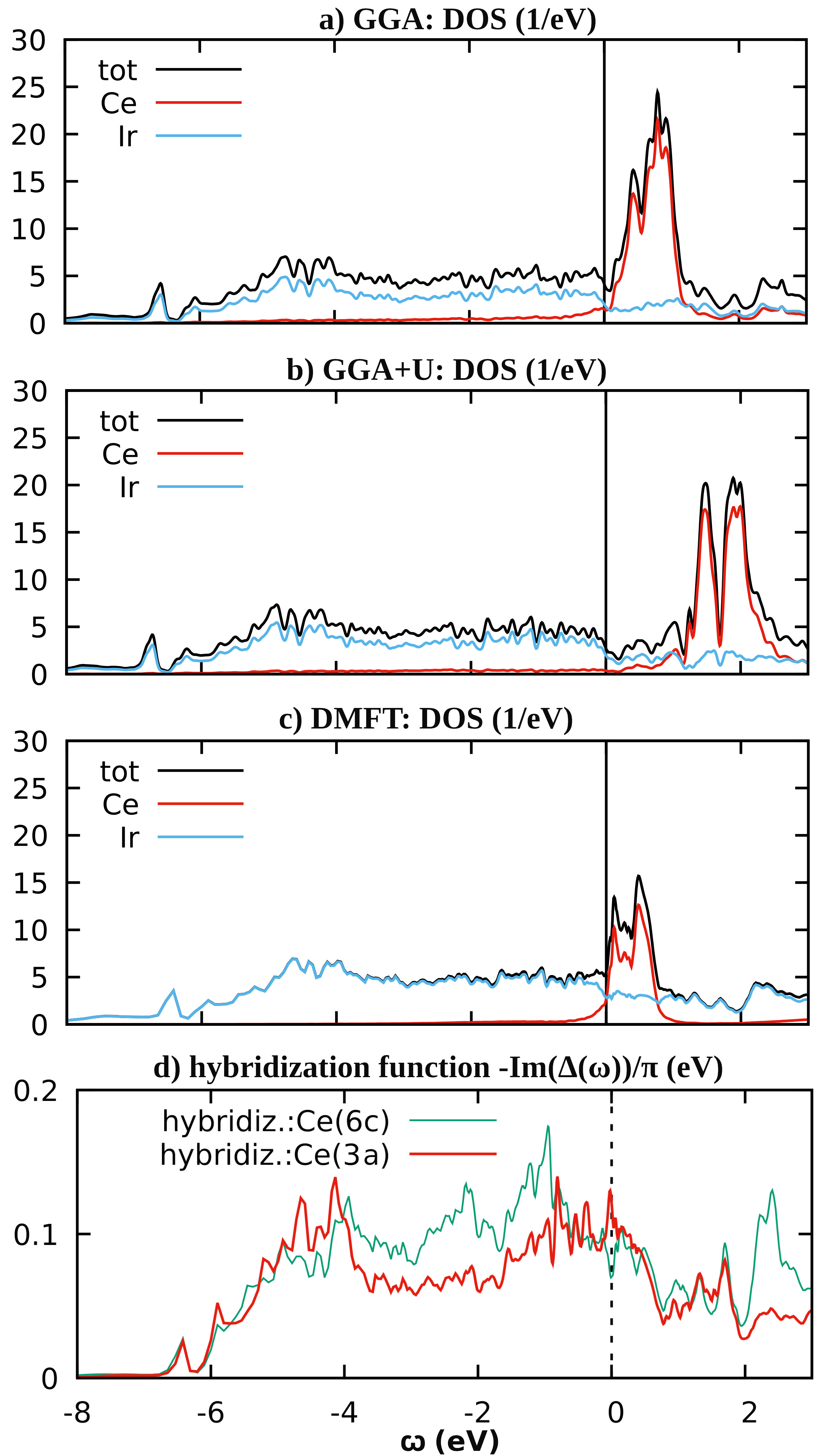}
    \caption{Comparison of DOS calculated with GGA  (a), GGA+U (b) and DMFT (c) methods, expressed in units of eV$^{-1}$ per f.u. Additionally the hybridization function of $4f$ states of two nonequivalent Ce atoms is presented in panel (d). All plots are shown with respect to Fermi energy.}
    \label{fig:dos}
\end{figure}

Densities of states (DOS) of CeIr$_3$ calculated with GGA and GGA+U methods (both including SOC) are shown in Fig. \ref{fig:dos}, for comparison of the Fermi surface and dispersion relations we refer the reader to Ref.~\cite{ceir3}. 
When GGA results are considered, the DOS at Fermi level is dominated by Ir:$5d$ states, however $4f$ states of Ce contribute significantly.

In Table \ref{tab:results} the properties of CeIr$_3$ calculated within GGA and GGA+U are summarized and compared to DMFT results, {discussed below}. The value of DOS at the Fermi level, obtained in the GGA calculations, is $N(E_F) =4.31$ eV$^{-1}$ per formula unit, and it gives the value of the Sommerfeld electronic specific heat parameter  
$\gamma_{\rm calc}= \frac{\pi}{3} k_B^2 N(E_F) = 10.16$\,$\frac{\rm{mJ}}{\rm{mol}\cdot\rm{K}^2}$.
As we mentioned, the experimental value of the Sommerfeld coefficient is equal to $\gamma_{\rm expt}=25.1$\,$\frac{\rm{mJ}}{\rm{mol}\cdot\rm{K}^2}$. The heat capacity renormalization coefficient, which in the electron-phonon superconductors is 
mostly contributed by the electron-phonon interaction, may be computed as
\begin{equation}
    \lambda_{\gamma}=\frac{\gamma_{\rm expt.}}{\gamma_{\rm calc.}}-1
    \label{lambda}
\end{equation}
and is equal to 1.47, in strong disagreement with the value of 0.65 estimated from the  McMillan's equation \cite{mcmillan}
\begin{equation}
\label{mcmillan}
 T_C=\frac{\theta_D}{1.45}\exp\left[\frac{-1.04(1+\lambda_{\rm ep})}{\lambda_{\rm ep} -\mu^*_0(1+0.62\lambda_{\rm ep})}\right]
 \end{equation} 
on the basis of critical temperature $T_C$, Debye temperature $\theta_D=142$\,K, with Coulomb pseudopotential parameter $\mu_0^*=0.13$. 
Any additional contribution to $\lambda_{\gamma}$ (e.g., from the electron-electron interaction) that would complete to such a large value would prevent from the formation of the superconducting phase. 
These discrepancies show that GGA method is not sufficient in the case of CeIr$_3$.
 Moreover, the DOS contributed by $4f$ state integrated up to the Fermi level leads to the large occupation of this state $n(4f)=1.05$.

When GGA+U approximation is used, with the Coulomb $U$ term applied to $4f$-states of Ce and assumed to be a constant $U=2.5$\,eV, the $4f$-states of Ce are pushed $1$\,eV above the Fermi level, as shown in Fig. \ref{fig:dos}(b). That leads to the smaller DOS at the Fermi level, $N(E_F)=2.42$ eV$^{-1}$, and smaller $\gamma_{\rm calc.}=9.20$\,$\frac{\rm{mJ}}{\rm{mol}\cdot\rm{K}^2}$. 
As a consequence, the situation with the electronic specific heat becomes even worse, as the renormalization parameter would have to reach $\lambda_{\gamma}=3.40$, unrealistic in the case of a rather typical electron-phonon superconductor, where strong electron-electron interactions are not expected.
The $4f$-state occupancy is small in this case, $n(4f)=0.2$. 
It is worth to note that in the presented calculations a rather weak Coulomb repulsion is assumed as $U=2.5$\,eV  is smaller than the value we have calculated with the help of the method developed by Madsen and Novak \cite{novak}.  The method, based on the formula for one-particle energy $\epsilon(n_i)=\epsilon_{LDA} +U(\frac{1}{2}-n_i)$, allows to calculate  $ U=\epsilon_i(n_i=0.5)-\epsilon_i(n_i=-0.5)$, where $\epsilon_i(n_i=\pm 0.5)$ is an energy of the system with half of $4f$ electron added (subtracted) to (from) the system and for CeIr$_3$ this method gives $U=6$\,eV, in agreement with the values commonly used for Ce. 
However, for the larger $U$, in CeIr$_3$ the discrepancy with the experimental electronic specific heat exacerbates: the $4f$ peak in DOS is pushed further above the Fermi level leading to a smaller $n(4f)$ and $N(E_F)$.
This shows that the GGA+U method, which usually improves the description of the electronic structure of materials with localized $4f$-states, fails in the case of CeIr$_3$, suggesting more complicated behavior of $4f$ electrons in this material.

\paragraph{eDMFT}

\begin{figure*}
    \centering
    \includegraphics[width=.99\textwidth]{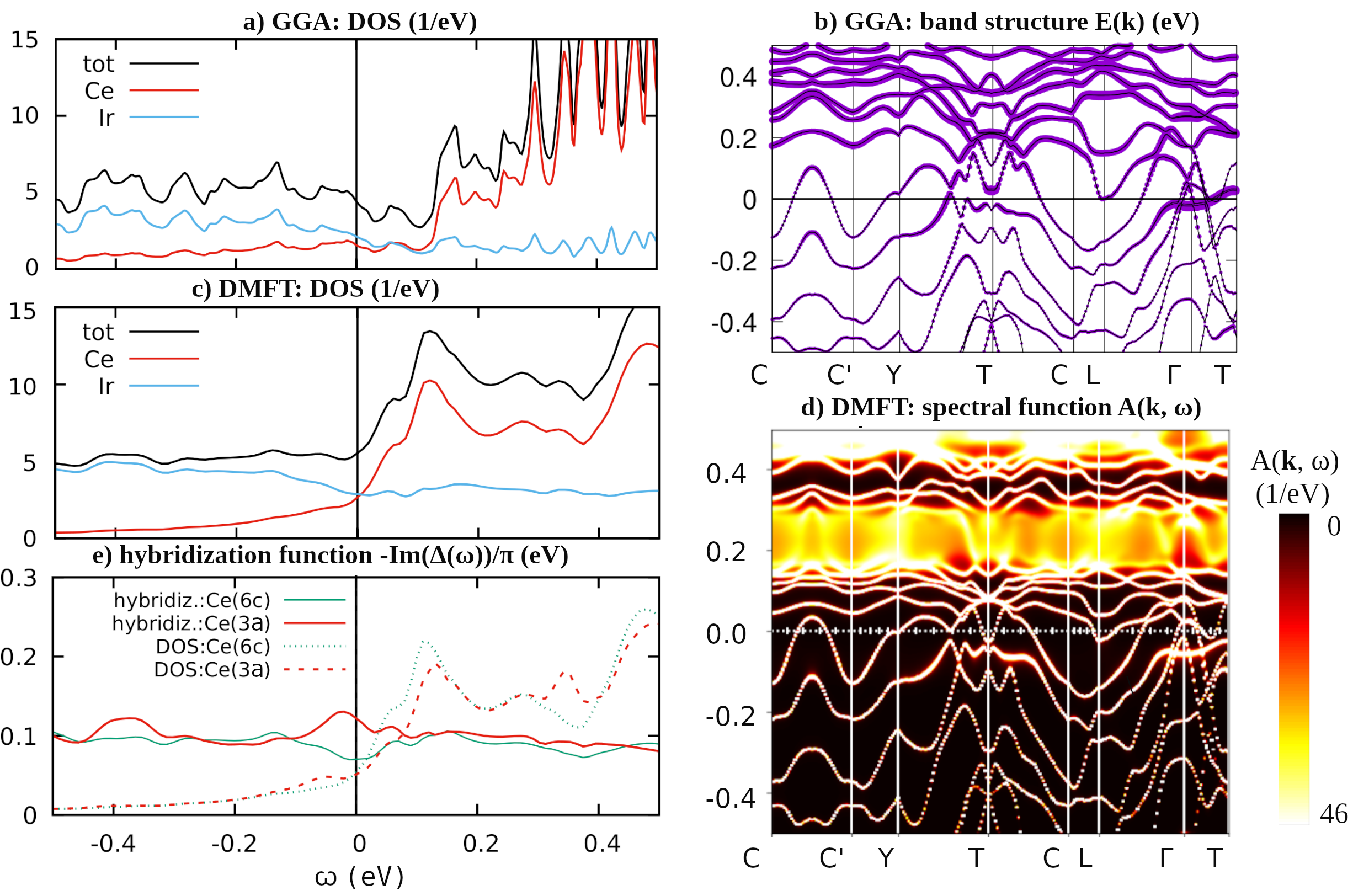}
    \caption{Comparison of band structure calculated with GGA  (a-b) and DMFT (c-e) methods, shown with respect to the Fermi energy. The GGA results are shown in terms of total and atomic DOS expressed in units of eV$^{-1}$ per f.u. (a) and dispersion relation with bandwidth proportional to contribution of $4f$ states (b). The DMFT results are presented by DOS (c), spectral function $A(\bf{k},\omega)$, where the linewidth is inversely proportional to lifetime of $4f$ state (d) and hybridization function of $4f$ states of two nonequivalent Ce atoms (see Fig. \ref{fig:cell}) together with partial DOS contributed by these states, divided by 50 (e).}
    \label{fig:bands}
\end{figure*}

Density of states of CeIr$_3$, calculated in the DMFT framework is shown in Fig. \ref{fig:dos}, where it is compared to GGA and GGA+U (all including the spin-orbit coupling).  
First, we see that the DMFT DOS is closer to GGA than to GGA+U, confirming that GGA+U fails in describing $4f$ electronic states in CeIr$_3$.
The shape of DOS calculated with the help of DMFT remains in agreement with GGA results deeply below the Fermi level, where Ir states dominate. 
Near $E_F$ in DMFT the hybridized $4f$ DOS is pushed to lower energy, increasing $N(E_F)$ value (see also Fig.~\ref{fig:bands} with zoom near the Fermi level). 
The hybridization function (precisely, the opposite of the imaginary part of $\Delta(\omega)$ defined in \cite{dmft2}) is shown in Fig. \ref{fig:dos}(d) and \ref{fig:bands}(e) near $E_F$. 
We can notice a difference between the two nonequivalent positions of Ce: Ce(3a), which builds the Ce-Ir layers in the crystal structure [see Fig. 3(e)], has a larger hybridization function at the Fermi level than Ce(6c). The latter is located between the layers, and its hybridization function is larger below -1\,eV.
This is followed by a larger contribution to DOS just below $E_F$ from the Ce(3a) rather than from Ce(6c) [see Fig. 3(e)], pointing to the importance of the layered type of crystal structure in the electronic properties of this material.
At the end, as shown in Tab.~\ref{tab:results}, the hybridization increases not only $N(E_F)$ contributed by Ce, but also that coming from Ir.

The probability of occupancy of $4f$ states is  calculated in the DMFT framework as a cumulative probability of $4f^n$ (n=1, 2, 3) atomic states of the total number of 106 atomic states considered in the 22 blocks. In case of CeIr$_3$, it is equal to 0.27 for $4f^0$, 0.67 for $4f^1$ and 0.06 for $4f^2$ state.
The occupancy of $4f^1$ state extracted from XPS measurements at room temperature for most of Ce$M_3$ compounds \cite{xps} varies from 0.64 to 0.80. For CeCo$_3$, the only compound of this series isostructural with CeIr$_3$, is $n(4f^1)=0.64$ (XAS measurements in \cite{xps-ceco3}), very close to that calculated here for CeIr$_3$.
The XPS measurements of CeIr$_3$ were presented in~\cite{ceir3}, and the measured relative intensity of the $4f^1$ peak, $I(f^1)/(I(f^0)+I(f^1)+I(f^2))$, was equal to 0.62, close to the calculated occupation of the $4f$ states.
Large hybridization energy $E_{hybrid}=0.2$\,eV was also deduced. 
Magnetic susceptibility ~\cite{ceir3} showed a weak temperature dependence and was analyzed using the  
interconfiguration fluctuation model (ICF) of intermediate valency \cite{icf}, in which $4f^1$ excited state is assumed to be magnetic. That led to a conclusion that mostly (in 95\%) the nonmagnetic $4f^0$ state is occupied. 
However, such a model seems to be insufficient in our case, where in DMFT calculations, due to the strong hybridization of the $4f$ levels, a nonmagnetic state is realized with a larger probability of $4f^1$ occupation.

Strong hybridization is followed by a relatively small effective mass,  calculated as a slope of the imaginary part of self energy $\frac{m^*}{m_e}=1-\Im(E)(\omega=0)$. Obtained $m^* = 1.7m_e$, which means, that bands are narrower than in the non-interacting picture, but the effect is weaker than in $\alpha-$Ce  ($m^*=5m_e$) \cite{edmft}, where $4f$ states are more localized.

Electronic dispersion relations of CeIr$_3$, calculated with GGA are compared to the spectral functions $A({\bf k},\omega)$ from DMFT in Fig. \ref{fig:bands}. In panel (b) the contribution from $4f$ states to the band structure is marked with a fatband, while in the spectral function (d) the bandwidth is inversely proportional to the electronic lifetime. 
Near $E_F$ and below, the shape of the bands are similar, we observe some shifts and changes in the relative position of the band maxima. Importantly, the DMFT spectral functions are narrow and a sharp well-defined band structure is observed, with a long electronic lifetime and delocalized electronic states.
In contrast, in the energy range from 0.1 to 0.3\,eV and above 0.4\,eV one can notice a strong band smearing due to enhanced electronic interactions and in these energy ranges strong electronic correlations appear.

Now coming to the analysis of the Sommerfeld parameter in DMFT, thanks to the strong hybridization of $4f$ electronic states of Ce with Ir orbitals near $E_F$ both Ir and Ce DOS at Fermi level increase (see Table~\ref{tab:results}), giving the total $N(E_F) = 5.83$\,eV$^{-1}$. When comparing the contribution to DOS at the Fermi level,  the DMFT shows that the hybridized $4f$ states contribute more to $N(E_F$) than $d$ states of Ir, if counted per atom. However, due to the larger population of Ir atoms, the overall contribution from three Ir atoms is 20\% larger than from Ce. 
As a consequence, the Sommerfeld coefficient increases to $\gamma_{\rm calc.}=13.74$\,$\frac{\rm{mJ}}{\rm{mol}\cdot\rm{K}^2}$. 
Now the specific heat renormalization parameter becomes significantly lower, $\lambda_{\gamma} = 0.83$, much closer to the electron-phonon coupling parameter estimated from the McMillan formula ($\lambda_{ep}(T_c) = 0.65$), than the GGA ($\lambda_{\gamma} =1.47$) or the GGA+U ($\lambda_{\gamma} =3.4$) results. 
This confirms the better accuracy of the DMFT description of the electronic structure of CeIr$_3$.

The analysis of the Sommerfeld parameter and the electron-phonon coupling may be improved if the presence of spin fluctuations, which were 
recently suggested to exist in CeIr$_3$ based on the $\mu$SR measurements \cite{ceir3-musr}, is taken into account.
Spin fluctuations, characterized by a coupling parameter $\lambda_{\rm sf}$, additionally renormalize the electronic heat capacity over the band-structure value~\cite{lambda_sf2}
$\gamma_{\rm expt.}=(\lambda_{\rm ep}+\lambda_{\rm sf}+1)\gamma_{\rm calc.}$.
On the other hand, they compete with superconductivity, renormalizing the electron-phonon coupling and Coulomb pseudopotential parameters. This effect may be approximately taken into account by using the McMillan formula, (Eq. \ref{mcmillan}), with $\lambda_{\rm ep.},~\mu^*_0$ replaced with the effective parameters~\cite{lambda_sf,lambda_sf2}
$\lambda_{\rm eff.} = \frac{\lambda_{\rm ep}}{1+\lambda_{\rm sf}}$
and 
$\mu^*_{\rm eff}=\frac{\mu^*_0+\lambda_{\rm sf}}{1+\lambda_{\rm sf}}$.
Taking the experimental value of $T_c$ and $\theta_D$, with typical $\mu^*_0 = 0.13$ we arrive at the small spin fluctuation parameter of $\lambda_{\rm sf}=0.05$ and electron-phonon coupling constant of $\lambda_{ep}=0.77$.
Similar situation, with $\lambda_{\rm sf}=0.10$ and  $\lambda_{ep}=0.74$, was recently found for 
ThIr$_3$ \cite{thir3}, which is isostructural to CeIr$_3$ and superconducts with $T_c=4.41$\,K.

\section{Summary}
The DMFT study of the electronic structure of CeIr$_3$ superconductor was presented. 
While GGA+U method treats $4f$ states as localized and pushes them above the Fermi level, DMFT show strong hybridization of $4f$ states, which causes their presence around the Fermi level, increasing the $N(E_F)$ value. 
The hybridization is connected to the layered type of crystal structure of CeIr$_3$ and is stronger for Ce atoms which form Ce-Ir planes.
The calculated occupation of $4f$ states is equal to 0.67, similarly to that observed in other Ce$M_3$ compounds. 
The Sommerfeld parameter and the electron-phonon coupling constant calculated on the basis of DMFT  
stands in good agreement with the experimental findings, in contrast to the previous GGA or GGA+U results. 
Furthermore, our analysis revealed weak spin fluctuations in agreement with the experimental data.

\section*{Acknowledgements}
This work was supported by the National Science Centre (Poland), project No. 2017/26/E/ST3/00119 and partly by the PL-Grid infrastructure.
S.G. was partly supported by the EU Project POWR.03.02.00-00-I004/16.

\bibliography{ceir3-refs}

\end{document}